\documentclass[conference]{IEEEtran}
\IEEEoverridecommandlockouts
\usepackage{cite}
\usepackage{url}

\usepackage{amsmath,amssymb,amsfonts}
\usepackage{amsthm}
\usepackage{algorithmic}
\usepackage{graphicx}
\usepackage[table]{xcolor}
\usepackage{subcaption}
\usepackage{textcomp}
\usepackage{enumitem}
\usepackage{listings}
\usepackage{multicol, multirow}
\usepackage{tikz}
\usepackage[switch]{lineno}
\usepackage{comment}
\usepackage{balance}

\usetikzlibrary{decorations.pathreplacing,calligraphy,fit}
\usepackage{xcolor}
\def\BibTeX{{\rm B\kern-.05em{\sc i\kern-.025em b}\kern-.08em
    T\kern-.1667em\lower.7ex\hbox{E}\kern-.125emX}}
\begin{document}

\title{Unveiling and Vanquishing Goroutine Leaks in Enterprise Microservices: A  Dynamic Analysis Approach}

\author{
\IEEEauthorblockN{Georgian-Vlad Saioc}
\textit{$\begin{array}{c}
     \textrm{Aarhus University, Uber Technologies, Inc.} \\
\textrm{Aarhus, Denmark} \\
\textrm{gvsaioc$@$cs.au.dk}
\end{array}$}
\and
\IEEEauthorblockN{Dmitriy Shirchenko}
\textit{$\begin{array}{c}
\textrm{Uber Technologies, Inc.} \\
\textrm{San Francisco, USA} \\
\textrm{shirchen$@$uber.com}
\end{array}$}
\and
\IEEEauthorblockN{Milind Chabbi}
\textit{$\begin{array}{c}
\textrm{Uber Technologies, Inc.} \\
\textrm{Sunnyvale, USA} \\
\textrm{milind$@$uber.com}
\end{array}$}
}

\newtheoremstyle{mystyle}
  {}
  {}
  {\itshape}
  {}
  {\bfseries}
  {.}
  { }
  {}

\theoremstyle{mystyle}

\newcommand{\numSvcPrecise}{2567}
\newcommand{\numSvc}{$\approx$2500}
\newcommand{\goSrcFiles}{$\approx$260K}
\newcommand{\goSrcFilesPrecise}{389260}
\newcommand{\goTestFiles}{142K}
\newcommand{\goTestFilesPrecise}{142855}
\newcommand{\goLOC}{$\approx$75 Million}
\newcommand{\goLOCPrecise}{76002355}
\newcommand{\goleakFound}{857}
\newcommand{\goleakPrevented}{260}
\newcommand{\goleakFoundInternal}{711 FIXME}
\newcommand{\leakproFound}{33}
\newcommand{\leakproFixed}{21}
\newcommand{\leakproSpeedup}{34\%}
\newcommand{\leakproMemSave}{9.2$\times$}
\newtheorem{fact}{Fact}
\newtheorem{corollary}{Corollary}
\newcommand{\numInstances}{200K}

\newcommand{\goTestTgt}{75K}
\newcommand{\goTestTgtPrecise}{77266}
\newcommand{\goAllLibTgt}{175K}
\newcommand{\goAllLibTgtPrecise}{177723}
\newcommand{\goSrcLibTgt}{135K}
\newcommand{\goSrcLibTgtPrecise}{136782}
\newcommand{\goNumTests}{450K}
\newcommand{\goNumTestsPrecise}{465804}

\newcommand{\gomela}{\textsc{Gomela}}
\newcommand{\goat}{\textsc{Goat}}
\newcommand{\gcatch}{\textsc{GCatch}}
\newcommand{\goleak}{\textsc{Goleak}}
\newcommand{\leakprof}{\textsc{LeakProf}}
\newcommand\todo[1]{\textcolor{red}{#1}}
\newcommand{\pd}{partial deadlock}
\newcommand{\pds}{partial deadlocks}
\newcommand{\PD}{Partial deadlock}
\newcommand{\PDs}{Partial deadlocks}

\newcommand{\go}{Go}
\newcommand{\company}{Uber}
\newcommand{\milind}[1]{{\color{red} milind: #1}}
\newcommand{\vlad}[1]{{\color{orange} Vlad: #1}}
\newtheorem{Definition}{Definition}[section]
\newtheorem{Theorem}{Theorem}
\newtheorem{Remark}{Remark}
\newtheorem{Observation}{Observation}
\newtheorem{Rule}{Rule}
\newtheorem{Fact}{Fact}
\newtheorem{Lemma}{Lemma}
\newtheorem{Corollary}{Corollary}
\newtheorem{Lcorol}{Corollary}
\newtheorem{Example}{Example}

\definecolor{dkgreen}{rgb}{0,0.6,0}
\definecolor{gray}{rgb}{0.5,0.5,0.5}
\definecolor{mauve}{rgb}{0.58,0,0.82}


\lstset{ %
aboveskip=5pt,
belowskip=0pt,
lineskip= {-1.5pt},
language=go,                %
basicstyle=\scriptsize,       %
numbers=left,                   %
numberstyle=\tiny,      %
stepnumber=1,                   %
numbersep=2pt,                  %
backgroundcolor=\color{white},  %
showspaces=false,               %
stringstyle=\scriptsize,
identifierstyle=\scriptsize,
commentstyle=\scriptsize,
basicstyle=\scriptsize\ttfamily,
showstringspaces=false,         %
showtabs=false,                 %
frame=tb,                   %
tabsize=2,                      %
captionpos=b,                   %
breaklines=true,                %
breakatwhitespace=false,        %
title=\lstname,                 %
keywordstyle=\color{red}\underbar,                                
escapechar={@},
}

\newcommand{\mgo}[2][\normalsize{}] {$\textrm{\lstinline[identifierstyle=#1\ttfamily{},basicstyle=#1\ttfamily{}]{#2}}$}

\tikzset{
    park/.style={
        rectangle,
        draw=black,
        fill=red!20,
        text width=12em,
        text=black,
        font={\footnotesize},
        minimum height=1.5em,
        text centered
    },
    chanop/.style={
        rectangle,
        draw=black,
        fill=blue!20,
        text width=12em,
        text=black,
        font={\footnotesize},
        minimum height=1.5em,
        text centered
    },
    caller/.style={
        rectangle,
        draw=black,
        fill=green!20,
        text width=12em,
        text=black,
        font={\footnotesize},
        minimum height=1.5em,
        text centered
    },
    stack/.style={
        rectangle,
        draw=green!20,
        minimum height=3em,
        fill=green!20,
        very thick,
        text width=7em,
        text=black,
        text centered,
    },
    data/.style={
        rectangle,
        draw=black,
        minimum height=3em,
        fill=white,
        very thick,
        text width=7em,
        text centered,
    },
    interpreter/.style={
        rectangle,
        draw=black,
        minimum height=3em,
        fill=black,
        text=white,
        very thick,
        text width=7em,
        text centered,
    },
    arrow/.style={->,>=stealth, thick},
    arrow-left/.style={<-,>=stealth}
}

\maketitle

\begin{abstract}
Go is a modern programming language gaining popularity in enterprise microservice systems. 
Concurrency is a first-class citizen in Go with lightweight ``goroutines'' as the building blocks of concurrent execution.
Go advocates message-passing to communicate and synchronize among goroutines.
Improper use of message passing in Go can result in ``\pd{}'' (interchangeably called a \textit{goroutine leak}), a subtle concurrency bug where a blocked sender (receiver) never finds a corresponding receiver (sender), causing the blocked goroutine to leak memory, via its call stack and objects reachable from the stack.

In this paper, we systematically study the prevalence of message passing and the resulting \pds{} in \goLOC{} lines of Uber's Go monorepo hosting \numSvc{} microservices.
We develop two lightweight, dynamic analysis tools: \goleak{} and \leakprof{}, designed to identify \pds{}.
\goleak{} detects \pds{} during unit testing and prevents the introduction of new bugs.
Conversely, \leakprof{} uses goroutine profiles obtained from services deployed in production to pinpoint intricate bugs arising from complex control flow, unexplored interleavings, or the absence of test coverage.
We share our experience and insights deploying these tools in developer workflows in a large industrial setting.
Using \goleak{} we unearthed \goleakFound{} pre-existing goroutine leaks in the legacy code and prevented the introduction of $\approx$\goleakPrevented{} new leaks over one year period.
Using \leakprof{} we found 24 and fixed 21 goroutine leaks, which resulted in up to \leakproSpeedup{} speedup and \leakproMemSave{} memory reduction in some of our production services.

\end{abstract}


\section{Introduction}
Go is a modern programming language gaining popularity in large-scale enterprise microservice~\cite{amazonWWW, netflixWWW, gitWWW, NadareishviliBook,FerdmanCloud,  MarioMicroservice, uberMicroWWW, zhang2022crisp, zhang2021optimistic}  systems. 
Go is among the world’s most popular programming languages, featuring consistently in the Tiobe index~\cite{tiobe}.
Its popularity in the industry is due to its simplicity and strength in the right feature set, including built-in concurrency, garbage collection, static typing, and good performance. 
Go has a vibrant software developer community with rich developer tools, frameworks, and libraries~\cite{awesomego}.

Concurrency is a first-class citizen in Go~\cite{tu2019understanding, zhang2021optimistic, chabbiGoDR}.
Any function invocation can be prefixed with the \mgo{go} keyword to run it asynchronously in the same process address space.
Such threads, dubbed \textit{goroutines}, are produced and managed by the Go runtime and, therefore, lightweight compared to OS threads.
An individual goroutine may create any number of other goroutines, but the lexical scope does not manage the lifetime of these child goroutines.
In other words, Go does not enforce any structured parallel programming (e.g., fork-join~\cite{DBLP:journals/annals/NymanL16}), which may result in child goroutines outliving their parents.
Since goroutine creation and scheduling are comparatively cheap, \go{} developers create them in abundance. 
For example, the profiling of about \numInstances{} production processes running in the data centers at \company{} showed that the median number of goroutines per process was $\approx\kern-0.25em2000$; in comparison, the number of threads for Java programs was around 256.

\begin{figure}[!t]
    \centering
    \includegraphics[width=\linewidth]{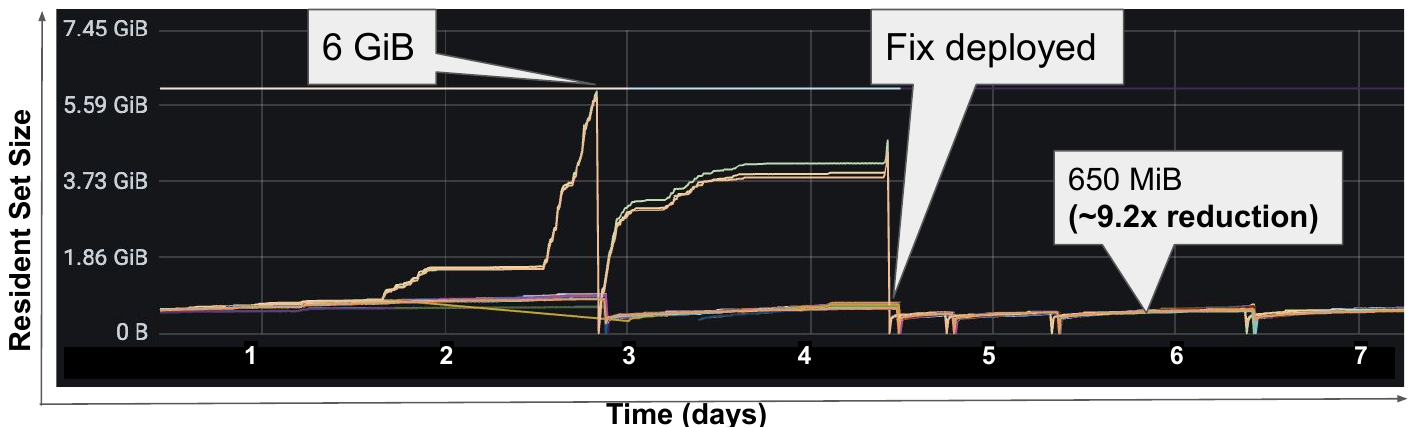}
    \caption{Resident set size (RSS) of a production microservice before and after fixing a  \pd{}. The fix reduces the RSS by 9.2$\times$. Different lines show the data points for different instances of the same program.}
    \label{fig:intoMemRed}
\end{figure}
Goroutines may communicate and synchronize via shared memory or message passing (despite being in the same address space).
Both programming paradigms are commonly used in Go with no particular winner.
The designers of Go, however, advocate~\cite{gomotto} for message-passing as the preferred synchronization mechanism.
In the message-passing model, two or more goroutines can communicate to send and receive messages over a shared \textit{channel}.
For \textit{synchronous} communication, the  goroutine performing a send operation blocks until a receiver goroutine issues a corresponding receive operation (or vice-versa) on the same \textit{unbuffered} channel, forming a rendezvous analogous to Hoare's Communicating Sequential Processes (CSP~\cite{DBLP:journals/cacm/Hoare78}).
Conversely, for \textit{asynchronous} communication involving \textit{buffered} channels, the sender blocks only if the channel buffer is full; conversely, receiving from a buffered channel blocks only if the buffer is empty or the channel is closed.

\begin{figure}[!t]
    \centering
    \includegraphics[width=\linewidth]{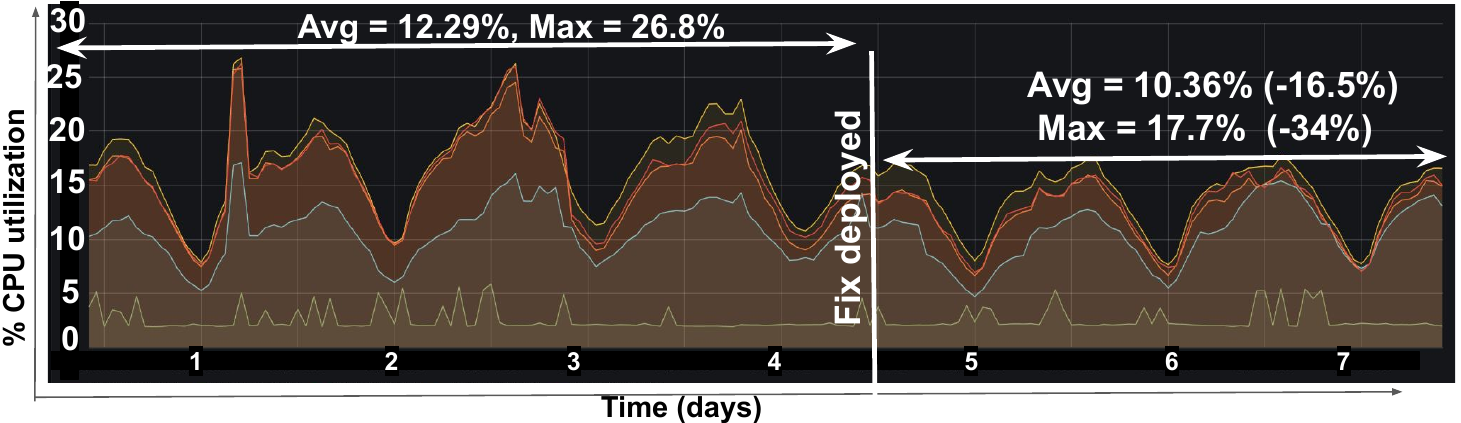}
    \caption{CPU consumption of a production microservice before and after fixing a  \pd{}. The fix reduces the max (average) CPU utilization by 34\% (16.5\%).  Different lines show the data points for different instances of the same program. Diurnal crests and troughs are common.}
    \label{fig:intoCPURed}
\end{figure}

Misuse of message passing can cause correctness defects such as deadlocks~\cite{luecke2002deadlock, MitchellDeadlock, tai1999deadlock}.
Correctness bugs arising from message passing represent only one kind of defect. 
Misuse of message passing in Go, however, can result in another kind of subtle bug called a \pd{} (Section \ref{sec:background}); \pds{} need not always manifest as functional defects and are thus harder to observe. 
\PDs{}, however, cause goroutines to leak, preventing them (and the objects they point to) from being garbage collected.
Eliminating goroutine \pds{} is essential for reliability and cost of capacity (computational resources) in industrial systems. 
Leaking goroutines can trigger sporadic out-of-memory problems, undermining the reliability of mission-critical systems. 
Furthermore, these leaks contribute to elevated memory consumption, driving up costs by necessitating greater DRAM capacity.

Fig~\ref{fig:intoMemRed} shows that leaking goroutines can squander as much as 9.2$\times$ extra memory, and 
Fig~\ref{fig:intoCPURed} shows that they can consume up to 34\% additional CPU time.
These aspects motivate our efforts to develop effective tools to detect \pds{} in \go{}, as presented in this paper.

We developed and deployed two \emph{dynamic} analysis tools---\goleak{} and \leakprof{}---to detect \pds{} in Go programs.
\goleak{} detects goroutine leaks by exercising unit tests; it is suitable for blocking any pull request (PR) containing new goroutine leaks that can be exposed via unit tests.
\leakprof{} monitors executions in production to detect goroutine leaks and automatically reports them when found; this is designed to detect goroutine leaks that may not be exhibited in unit testing but may manifest in complex control flows and thread interleavings that happen in production.
The two tools combine to prevent and detect goroutine leaks in production systems.

We have deployed these tools at \company{}'s \goLOC{} lines of Go monorepo code, which hosts \numSvc{} Go microservices, which run on millions of CPU cores.
\goleak{} detected \goleakFound{} unique goroutine leaks in the pre-existing code, 
and it has prevented an estimated \goleakPrevented{} new leaks from being introduced during its one-year of deployment.
During the same period, using \leakprof{}, we found 24 and fixed 21 goroutine leaks, resulting in as much as \leakproSpeedup{} savings in CPU and \leakproMemSave{} memory reduction.

The contributions of this paper are the following:
\begin{enumerate}
    \item We develop two complementary lightweight dynamic analysis tools---\goleak{} and \leakprof{}---to detect goroutine leaks. \goleak{} detects goroutine leaks by exercising unit tests. \leakprof{}  detects goroutine leaks in production executions.
    \item We describe our experience deploying these tools on a large industrial Go monorepo.
    \item Using \goleak{}, we found \goleakFound{} leaks and prevented an estimated  \goleakPrevented{} new leaks.    
    \item Using \leakprof{} we found 24 leaks, of which 21 were fixed, which resulted in a reduction in memory consumption of services by as much as \leakproMemSave{} and a reduction in CPU consumption by \leakproSpeedup{}.   
\end{enumerate}

The rest of the paper is organized as follows:
Section~\ref{sec:background} provides the background on \pds{}, features of Go, and state-of-the-art geared towards \pd{} detection;
Section~\ref{sec:methodology} explains our methodology and the motivation to pivot to dynamic analysis;
Section~\ref{sec:goleak-internals} and Section~\ref{sec:leakprof-internals} describe the designs of \goleak{} and \leakprof{}, respectively;
Section~\ref{sec:goleak_findings} and Section~\ref{sec:lekprof_findings} showcase the findings from production deployment of \goleak{} and \leakprof{}, respectively;
Section~\ref{sec:conclusion} offers our conclusions and future directions.

\section{Background and Motivation}
\label{sec:background}
\begin{figure}[!t]
\begin{lstlisting}[language=Go,label={lst:example},caption=A sample goroutine \pd{}.]
// File path: .../transactions/cost.go
package transactions

func (s *Server) ComputeCost(item *Item) (Cost, error) {
@\label{lst:ex:mkch}@  ch := make(chan *Discount) // Discount channel
@\label{lst:ex:mkgo}@  go func() {
@\label{lst:ex:fetchdisc}@    disc := s.getDiscount(item)
@\label{lst:ex:sendmsg}@    ch <- disc // Send discount over ch
  }()

@\label{lst:ex:fetchcost}@  amount, err := s.getBaseCost(item)
  if err != nil {
@\label{lst:ex:ret}@    return nil, err
  }
@\label{lst:ex:recv}@  disc := <-ch // Receive discount from ch
  return amount.applyDiscount(disc), nil  
}
\end{lstlisting}
\end{figure}

Listing~\ref{lst:example} shows an example of channel communication in Go.
The parent function \mgo{ComputeCost} has two concurrent operations
\mgo{getDiscount} and \mgo{getBaseCost}.
It creates an unbuffered channel \mgo{ch} on Line~\ref{lst:ex:mkch}, runs \mgo{getDiscount} asynchronously on Line~\ref{lst:ex:mkgo}, and then calls \mgo{getBaseCost} on Line~\ref{lst:ex:fetchcost}.
If \mgo{getBaseCost} succeeds (\mgo{err == nil)}, it then waits to receive (denoted as \mgo{<-} to the left of the channel) the discount on Line~\ref{lst:ex:recv}.
The \emph{anonymous} closure launched as a goroutine on Line~\ref{lst:ex:mkgo} sends (denoted as \mgo{<-} to the right of the channel) the discount on Line~\ref{lst:ex:sendmsg}.
This channel-send operation blocks and waits until the corresponding receive on Line~\ref{lst:ex:recv} is ready, and vice-versa.
However, in the abnormal case when \mgo{getBaseCost} on Line~\ref{lst:ex:fetchcost} returns an error, the parent goroutine \mgo{ComputeCost} returns prematurely on Line~\ref{lst:ex:ret} without attempting to receive the message from the channel.
In this situation, the child goroutine blocks upon sending the message at Line~\ref{lst:ex:sendmsg} and waits forever since there is no longer a receiver to pair up with. 
The failure of a goroutine blocked on a channel operation to rendezvous is known as a \textit{\pd{}} or a \textit{goroutine leak}.
We use the term \pd{} to delineate it from the traditional definition of \textit{deadlock}~\cite{SilberschatzOS}, which requires a cyclic wait, non-preemption, and mutual exclusion.

\textbf{The Select statement:} Another Go feature closely related to channels is the select statement, which allows a goroutine to simultaneously wait on multiple channel operations using different \mgo{case} arms (see Listing~\ref{lst:select}), each potentially guarding a sequence of statements.
The first, ready channel operation and its corresponding case arm execute at runtime.
If multiple arms are simultaneously ready, the runtime picks one non-deterministically.
Select statements may also feature an optional \mgo{default} case arm, which executes if no other case is ready, making the select statement non-blocking.
A select statement with no cases blocks forever.

\begin{figure}[!t]
\begin{lstlisting}[language=Go,label={lst:select},caption=Example select statement.]
func MyFunc(ch1 chan int, ch2 chan float) {
  select {
    case v1 := <- ch1 // receive from ch1
    case ch2 <- 42.0 // send 42.0 to ch2
    default: // executes if both ch1 and ch2 are blocked
  }
}
\end{lstlisting}
\end{figure}

\textbf{Impact of \pd{}:} A goroutine stuck due to a \pd{} leaks memory because its stack and any heap resources reachable through its stack (including e.g., the blocking channel) cannot be garbage collected. 
Referring back to Listing~\ref{lst:example}, the stack space for the anonymous goroutine and also the heap objects \mgo{ch}, \mgo{s}, \mgo{item}, and \mgo{disc} leak.

Unlike a \textit{global} deadlock (e.g., in the OS kernel), where the entire system hangs, a \pd{} only compromises parts of the system. 
The design pattern of a typical microservice request typically involves splitting the workload into several sub-tasks, which may then be run in parallel.
The parent task establishes a communication channel with the children tasks to wait for their results.
Top-level tasks are generally written to be resilient to failures by incorporating fallback/retry/timeout mechanisms.
If the parent task terminates without waiting for a response from the subtasks, such subtasks can accumulate into \pds{}.

Furthermore, programs are deployed in large numbers over many hosts, dispersing \pds{} across many program instances.
The confluence of these factors may lead to substantial segments of compromised memory across the entire platform.

Unless the goroutine leak causes an out-of-memory situation quickly after starting the program, such leaks may go unnoticed for a while and steadily degrade application and system performance. 
Unfortunately, service owners may be unaware of such leaks because services get redeployed every few days in fast development cycles, eliding the leak. Alternatively, the underlying resource allocation framework may automatically scale the service e.g., by provisioning additional hosts, CPU, and memory based on consumption trends.
In the worst case, the developers sometimes address the growing memory consumption by merely over-provisioning service with higher memory without realizing it to be caused by a \pd{}.


\subsection{\company{}'s Go code characteristics}
\label{sec:go-charac}
\company{} employs a single repository (monorepo) for all Go services and internal tools.
It uses the Bazel~\cite{bazel} build system to build and test code, which relies on the following terminology:
\begin{itemize}
    \item  A \textit{target} is the smallest compilation unit, e.g., a library (known as a \textit{package} in Go).
    \item  Packages can be composed to form \textit{dependency chains}.
    \item A target is further qualified as a \textit{test target} if it consists of the test suite of another package.
\end{itemize}

Table~\ref{tab:go-features} shows some of the salient features of Go as used throughout the monorepo.
These include the spawning of goroutines, channel allocations, and operations (send/receive operations). 

\begin{table}[!t]
\scriptsize
    \centering
    \begin{tabular}{|r||c|c|c|c|}
\hline
\textbf{Concurrency paradigm} & \textbf{Packages} & \multicolumn{2}{|c|}{\textbf{Files}} & \textbf{ELoC}\\
\hline
\multirow{2}{*}{\textbf{Message passing (MP)}} & \multirow{2}{*}{4,699}  & \textbf{Source} & 22K & 3.39M\\
\cline{3-5}
 && \textbf{Tests} & 15K & 4.81M \\
\hline
\multirow{2}{*}{\textbf{Shared Memory (SM)}} & \multirow{2}{*}{6,627} & \textbf{Source} & 29K & 4.87M\\
\cline{3-5}
&& \textbf{Tests} & 20K & 6.17M\\
\hline
\multirow{2}{*}{\textbf{MP $\cap$ SM}} & \multirow{2}{*}{2,416} & \textbf{Source} & 13K & 2.28M \\
\cline{3-5}
& & \textbf{Tests} & 10K & 3.26M \\
\hline
\hline
\multirow{2}{*}{\textbf{Entire monorepo}} & \multirow{2}{*}{119,816} &\textbf{Source}  & 260K & 46.31M
\\
\cline{3-5}
&& \textbf{Tests} & 142K & 29.37M
\\
\hline
\end{tabular}
    \caption{Distribution of Go packages with concurrency features, including the number of files and effective lines of code (ELoC) at \company{}'s Go monorepo.}
    \label{tab:go-packages}
\end{table}

\begin{table}[!t]
\small
\centering
\begin{tabular}{|r|c|c|}
\hline
\textbf{Feature} & \textbf{Source} & 	\textbf{Tests} \\
\hline
\hline
\multicolumn{3}{|c|}{\cellcolor{blue!25}\textbf{Functions}} \\
\hline
Anonymous & 31,000 & 41,785 \\
\hline
Named & 1,025,687 & 32,666 \\
\hline
With channel parameter(0 & 2,410 & 565 \\
\hline
With channel return type(s) & 1,387 & 1,387  \\
\hline
\hline
\multicolumn{3}{|c|}{\cellcolor{blue!25}\textbf{Goroutine creation}}\\
\hline
Via \mgo[\scriptsize]{go} keyword & 11,136 & 3,745 \\
\hline
Via wrapper function & 5,342 & 366 \\
\hline
\textbf{Total} & 16,478 & 4,111 \\
\hline
\hline
\multicolumn{3}{|c|}{\cellcolor{blue!25}\textbf{Channel allocations via \mgo[\scriptsize]{make(chan)}}}\\
\hline
Unbuffered & 3,006 & 3,444 \\
\hline		
Size-1 buffers & 1,295 & 1,175 \\
\hline
Constant ($>$1) buffers & 328 & 435 \\
\hline
Dynamically sized  buffers	& 2,018 & 270 \\
\hline
\textbf{Total} & 6,647 & 5,324 \\
\hline
\hline
\multicolumn{3}{|c|}{\cellcolor{blue!25}\textbf{Channel operations}}\\
\hline
Sends: \mgo[\scriptsize]{c<-} & 7,803 & 3,440 \\
\hline
Receives: \mgo[\scriptsize]{<-c} & 9,584 & 6,586 \\
\hline
\mgo[\scriptsize]{close} & 4,078 & 2,117 \\
\hline
\hline
\multicolumn{3}{|c|}{\cellcolor{blue!25}\textbf{\mgo[\scriptsize]{select} statements}}\\
\hline
Blocking & 3,046 & 965 \\
\hline
Non-blocking & 1,052 & 430 \\
\hline
\textbf{Total} & 4,098 & 1,395 \\
\hline
\hline
\multicolumn{3}{|c|}{\cellcolor{blue!25}\textbf{Overall cases in blocking \mgo[\scriptsize]{select}}}\\
\hline
P50 (50$^{th}$ percentile) & 2 & 2 \\
\hline
P90 (90$^{th}$ percentile) & 3 & 2 \\
\hline
Maximum & 11 & 6 \\
\hline
Statistical mode (most common) & 2 & 2 \\
\hline
\end{tabular}
    \caption{Features of Go and their prominence in message passing packages local to the monorepo at \company{}.}
    \label{tab:go-features}
\end{table}

The key takeaways are as follows: 
\begin{enumerate}
    \item Goroutine creation is pervasive, which is also documented in a prior study~\cite{chabbiGoDR}. 
    \item Production code uses several wrappers to create goroutines, as opposed to simply using the \mgo{go} keyword.
    \item Channel synchronization and communication are common.
    \item Unbuffered channels are commonly used.
\end{enumerate}

\subsection{Prior art}
\label{sec:state-of-the-art}
The development of program analysis tools~\cite{DBLP:conf/asplos/LiuZQCS21, DBLP:conf/kbse/VeileborgSM22, DBLP:journals/corr/abs-2004-01323, DBLP:conf/sas/MidtgaardNN18} to detect \pds{} is an active research area. 
Prior efforts have primarily employed a static program analysis approach.
\gcatch{}~\cite{DBLP:conf/asplos/LiuZQCS21} employs a constraint solver.
\goat{}~\cite{DBLP:conf/kbse/VeileborgSM22} performs abstract interpretation.
\gomela~\cite{DBLP:conf/kbse/DilleyL21} converts parts of Go programs into Promela~\cite{HolzmannPromela} programs for model checking.

\gcatch{} and \goat{} work on the whole-program level and rely upon expensive points-to alias analysis~\cite{wwwgopointer,andersen1994program} to build the call graph and over-approximate the inter-procedural scope of channels.
On a large industrial monorepo with $\approx\kern-0.3em 1640$ external dependencies, the execution time of points-to analysis may take as long as three hours, making it cumbersome for deploying in continuous integration (CI) and continuous delivery (CD) systems, which has thousands of code commits per day.
Furthermore, the points-to may-alias pre-analysis severely over-approximates the call graph, causing the analysis to consider many spurious execution paths. 
Consequently, unsound but pragmatic heuristics are often used to improve precision and scalability e.g., both \goat{} and \gcatch{} use similar heuristics for grouping channels that should be analyzed together.
The lowest-common ancestor (LCA) in the call graph, covering the operations of all channels in the group, is then used as the analysis entry point. 
This removes the need to analyze callers of the LCA, callees that do not involve channels in the group, and operations of channels not in the group, which are considered non-blocking no-ops. 
Despite these best efforts at constraining the scope of the analysis, the significant degradation in precision introduced by the points-to analysis still leads to a large number of both false positives (noisy incorrectly classified \pds{}) and negatives (missing real defects) (Section \ref{sec:methodology}).

The differences between \goat{} and \gcatch{} lie in the underlying techniques.
\goat{} relies on abstract interpretation~\cite{DBLP:journals/csur/Cousot96}, constructing a least-fixpoint over conservative approximations of the program state. This alternates between issues with either precision or scaling. 
In general, better precision leads to better recall, but analysis of Go programs may become imprecise due to several features: aliasing, control flow, reflection, interleavings of concurrent operations or non-deterministic internal choices. 
Handling these features through known techniques naively typically leads to many spurious reports and may inadvertently increase the analysis time, so unsound but pragmatic heuristics are employed to reduce complexity and improve precision, e.g., by explicitly modeling certain standard library features.

In contrast, \gcatch{} enumerates bounded execution paths, which are then encoded as SMT formulae.
The derived formulae model ordering of operations in the program and the semantics of channel operations, namely blocking, non-blocking, and unsafe behavior. 
The formulae are fed to Z3~\cite{DBLP:conf/tacas/MouraB08}, an SMT solver, and any operation that is deemed reachable but unable to show progress is reported as a blocking error.
%
%

\gomela{} relies on model checking by translating Go programs to Promela models, which are then checked with SPIN\cite{DBLP:journals/tse/Holzmann97}. 
The symmetry in many syntactical features between the two languages makes intraprocedural translation convenient.
Another advantage of relegating the analysis to the AST is that the constructed Promela models preserve the structure of intra-procedural control flow.
Foregoing the points-to analysis also allows \gomela{} to operate without a program entry point\footnote{\goat{} and \gcatch{}  circumvent the absence of a \mgo[\footnotesize]{main} function in libraries by using tests as artificial program entry points. This inadvertently ties their scope to the test suite coverage.}, giving it the distinct advantage of analyzing code embedded deeply into libraries.
On the flip side, in its current state, its inter-procedural reasoning capabilities are limited to only pursuing anonymous functions that are called immediately or statically known call edges.
As a result, programs featuring higher-order functions, such as wrappers around \mgo{go} instructions or that involve dynamic dispatch, typically blindside it and require ad-hoc amendments to the existing technique. 
Another issue with \gomela{} and other tools that rely on model checking is that they may run out of memory during the model-checking phase or take too long. To circumvent this, we have imposed a 60-second per-model verification time limit.

Table \ref{tab:static-tools} shows the relative performance of various  analysis tools. The \textbf{Overhead} column shows both Offline (compile/test/analyze) and Online (runtime) overhead.
\textbf{Reports} column shows the number of alerts for each tool.
\textbf{Precision} is the ratio of true positives to the total number of alerts.
For the static analysis tools, we picked a random set of 114 reports per tool for manual inspection. 
While \goat{} and \gcatch{} have higher precision, at $47\%$ and $51\%$ respectively, due to their stronger inter-procedural and aliasing reasoning capabilities, they still introduce too many false positives, detracting from their usefulness in an industrial environment.
Furthermore, while the number of false negatives is unknown, the front-end limitations of each tool suggest that the upper bound of false negatives is potentially very high.
During an evaluation, we found all tools insufficiently mature to deploy in an industrial fast development setting (thousands of commits per day) with continuous integration (CI) without significant adjustments to their configurability. 
A shared drawback is that hiding concurrent operations behind high-level APIs e.g., spawning goroutines via wrappers, severely impedes the detection of \pds{} unless such API calls are properly recognized.
Extending the analyses with the capacity to recognize wrapper APIs as synchronization points is cumbersome, as the set keeps changing and growing.

\begin{table}[!t]
\scriptsize
    \centering
    \begin{tabular}{|c||c|c|c|c|}
    \hline
    \multirow{2}{*}{
    $\begin{array}{@{} c @{} }
        \textrm{\textbf{Tool}} 
    \end{array}$} &
    \multicolumn{1}{|c|}{
    \multirow{1}{*}{
    $\begin{array}{@{} c @{} }
        \textrm{\textbf{Overhead}}
    \end{array}$}} &
    \hspace{-0.8em}
    \multirow{2}{*}{
    $\begin{array}{@{} c @{} }
        \textrm{\textbf{Reports}}
    \end{array}$} 
    \hspace{-0.6em} &
    \multirow{2}{*}{
    $\begin{array}{@{} c @{} }
        \textrm{\textbf{Precision}}
    \end{array}$} &
    \hspace{-0.8em}
    \multirow{2}{*}{
    $\begin{array}{@{} c @{} }
        \textrm{\textbf{Deployable}}\\
        \textrm{\textbf{in CI/CD}}
    \end{array}$}
    \hspace{-0.6em} \\
    \cline{2-2}
    & \hspace{-0.6em} \textbf{Offline / Online} 
    \hspace{-0.6em} & & & \\
    \hline
    \multicolumn{5}{|c|}{\textbf{Static analyses}}\\
    \hline
    \gcatch{} & high$^*$ / None & 938 & 59/114 (51\%)  & No \\ \hline
    \goat{} & high$^*$ / None & 450 & 54/114 (47\%)  & No \\ \hline
    \gomela{} & high$^\#$ / None & 389 & 39/114 (34\%)  & No\\
    \hline
    \multicolumn{5}{|c|}{\textbf{Dynamic analyses}}\\
    \hline
    \goleak{} & low / None & \goleakFound{} & 100\% & Yes\\ 
    \hline
    \hspace{-0.6em}
    \leakprof{} 
    \hspace{-0.6em} & low / low & 33 & 72.7\%  & No$^+$\\
    \hline
    \end{tabular}
    \caption{Performance overview of analysis tools.\\
    \footnotesize{$^*$ Deploys SSA-IR translation and Andersen's points-to analysis}\\
    \footnotesize{$^\#$ Operates directly on the AST; leverages SPIN\cite{DBLP:journals/tse/Holzmann97} for model checking\\
    $^+$ Not designed for testing but can be retrofitted to work on tests. }}
    \label{tab:static-tools}
\end{table}

\section{Methodology}
\label{sec:methodology}

We desire the following characteristics for a deployable goroutine \pd{} detection tool:
\begin{description}
    \item [Lightweight:] The tool should produce results in a matter of seconds for each code change so that it can be integrated with CI/CD systems;
    \item [High precision:] $(true\ positives)/(true\ positives + false\ positives)$ should be high so that the developers are not inundated with noisy results; and
    \item [Deployability:] Good support for abstractions (e.g., wrappers, libraries) used in the large, existing code base.
\end{description}

\begin{figure*}[!t]
    \centering
    \includegraphics[width=.8\linewidth]{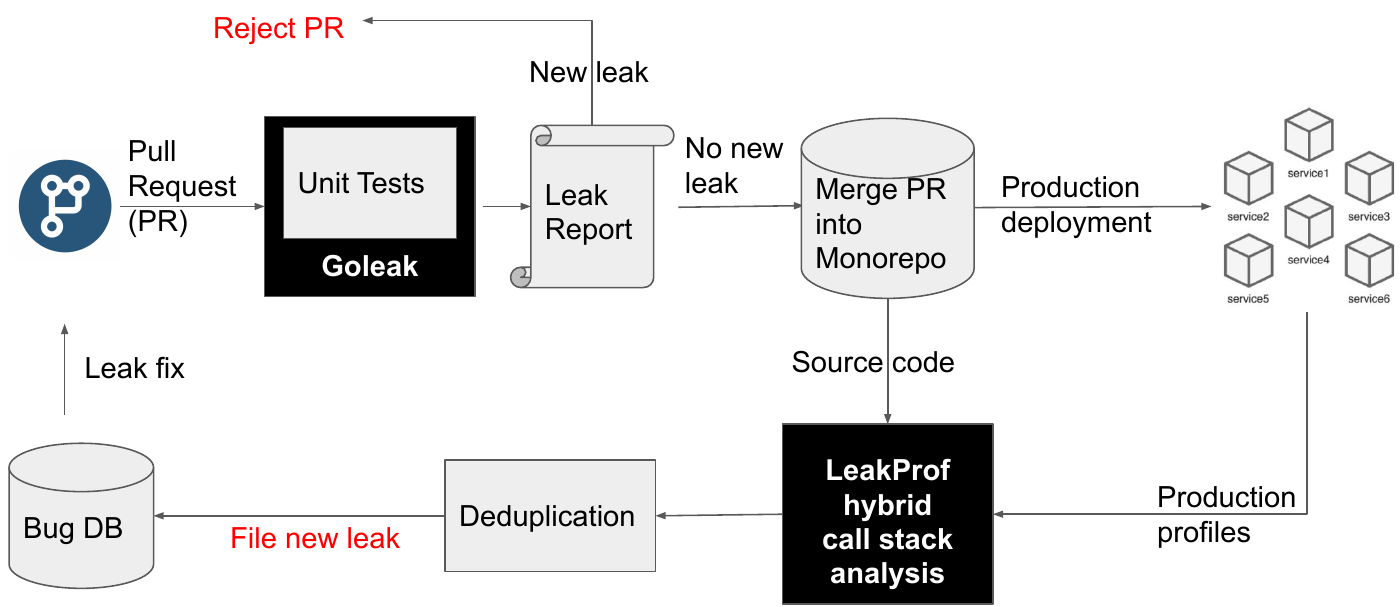}
    \caption{Schematic diagram of dynamic analysis for \pd{} detection in Go monorepo at \company{}. Unit tests are instrumented with \goleak{}, and PR is rejected if a new leak is found. Successful PRs are merged and deployed into production. \leakprof{} analyzes call stacks in production profiles alongside the source code and reports potential \pds{}.}
    \label{fig:schematic}
\end{figure*}

The aforementioned three static analysis tools fell short in all three dimensions; for that reason, we pivoted to lightweight dynamic analysis techniques and developed two complementary tools---\goleak{} and \leakprof{}---that scale to our systems.
Our methodology for detecting \pds{} on \company{}'s development and production settings is depicted in Fig~\ref{fig:schematic}. 
The first tool \goleak{} runs in our CI systems on every PR; it detects goroutine leaks by exercising unit tests and blocks any PR containing new goroutine leaks; with over \goNumTests{} unit tests in our repository, this serves as an effective technique to detect leaks early in the development cycle.
Code authors should fix the leaks before they are reviewed and merged into the main branch.
The second tool \leakprof{}, periodically profiles production executions, analyzes call stack samples, and automatically alerts service owners if it finds symptoms of goroutine leaks via a dynamic program analysis; this is designed to detect goroutine leaks that may not manifest in unit testing but may be revealed by the complex control flow and space of thread interleavings explored in long-running production executions.

Both tools are lightweight, show negligible overheads in production, have very high precision, and, being dynamic, neither needs additional support to handle abstractions.

\section{\goleak{}}
\label{sec:goleak-internals}

\subsection{Internals}
\goleak{} is the simpler of the two tools, yet effective in curbing partial deadlocks at their source---when new code is introduced.
In Go, testing is a built-in feature: files suffixed with \texttt{\_test.go} in a Go package mark test files, and functions prefixed with \mgo{Test} in such files are considered test units.
One can exercise all tests in a package via the \texttt{go test} command. 
The Go testing system internally creates a standalone executable that invokes each \mgo{Test} prefixed function.

\company{}'s Go code base has over \goNumTests{} tests composed in \goTestFiles{} test files organized in \goTestTgt{} test targets.
New PRs exercise tests in all packages affected by code changes, and any introduced failure blocks the PR.
This makes detecting \pds{} through the test suite and blocking PRs a promising approach.
Unit tests are written to exercise the correctness of some business logic. However, developers rarely target tests to expose the behavior of a channel used in their business logic code. 

\begin{fact}
If a goroutine is partially deadlocked, it will remain in the process address space until program termination.
\end{fact}

\begin{corollary}
If any goroutine is present at program termination, it may be partially deadlocked. 
\end{corollary}

\goleak{} is developed assuming that at the end of exercising a test, there shall be no lingering goroutines in the process address space. 
Naturally, every \pd{} will violate this assumption.
The converse, however, is not true---lingering goroutines at the end of a test need not necessarily be a \pd{}.
Other than blocking on a channel, the reasons for a goroutine to linger at the end of a test are: sleeping on a timer, waiting for an IO operation, and running computation, to name a few.
Nevertheless, the assumption is still a useful golden standard, motivating the developers to fix any lingering goroutines in exercised tests, regardless of the underlying cause.

We developed and deployed the \goleak{} dynamic analysis tool with this goal.
\goleak{} is a library providing APIs (such as \mgo{Find}) to collect the call stacks of all goroutines currently present in the process address space, excluding the caller itself. 
The data contains the following information for each goroutine\footnote{Internally, we rely on the Go runtime-provided \mgo{Stacks} API to gather such information.}:
\begin{description}    
\item[status:] which can be one of blocked on send/receive/select operations, running, or waiting for semaphore
\item[code context:] which is the function name along with its source location at the leaf of the goroutine call stack
\item[creation context:] which is the  function name source location where the goroutine was created
\end{description}

The first challenge in deploying \goleak{} is instrumenting tests to invoke it.
If instrumentation must be carried out manually, a developer, whether deliberately or not, may skip it, potentially voiding its utility. 
To reduce the friction of manual instrumentation, we developed tooling to automatically instrument tests as part of the build pipeline, such that \goleak{} is invoked at the end of the execution of each test target.
The special Go function, \mgo{TestMain}, is a hook that is executed upon running the entire test suite.
During test execution, we transparently patch test targets to introduce wholesale, or otherwise amend, \mgo{TestMain} with an invocation of \goleak{}.
The instrumentation invokes \mgo{goleak.VerifyTestMain}, a special \goleak{} API that first executes all tests in the target, logs stack traces if any lingering goroutines are found in the address space, and marks the test target as failed.

At a large industrial scale, deploying automated instrumentation poses an additional challenge, specifically its timing and handling.
Indiscriminate instrumentation of existing tests may cause a sudden influx of errors, blocking numerous PRs exercising tests with pre-existing goroutine leaks but not introducing any new leaks by themselves; this can slow down the development pipeline and potentially compromise timely delivery. 
To address this concern, we performed an offline trial run with instrumentation and collected all leaking goroutine locations as function names.
These were then added to a suppression list that prevents the blocking of PRs with test failures at these locations.
This allows test owners to gradually address leaks in the existing code base conveniently, and then correspondingly remove affected tests from the suppression list instead of suddenly being overwhelmed by all the previously unknown issues.
Only PRs that introduce new \pds{} may then be blocked.

\subsection{Goleak overhead} 
Running all \goNumTests{} tests with \goleak{} enabled showed statistically insignificant overhead; the tests take tens of hours of compute time.

To assess the worst-case overhead, we created a pathological test that only generated a large number of partially deadlocked goroutines but did nothing else in the test.
We measured the overhead to range from $4.6\times$-$7.4\times$.
The overhead increases with the increasing number of partially deadlocked because \goleak{} has to do more work to walk more call stacks and report them.
A single call stack unwind adds 200-400$\mu s$ overhead.

\section{\leakprof{}}
\label{sec:leakprof-internals}

Pprof\cite{pprof} is the standard profiling tool built into the Go runtime.
\leakprof{} relies on information it provides in the form of goroutine profiles from production instances, which it then analyzes at scale through lightweight techniques to detect goroutine leaks.

\subsection{Internals}
\label{sec:leakprof-profiles}
\textbf{Profile collection:}
Pprof allows for collecting an instantaneous snapshot of call stacks of all goroutines in the program, which is known as a goroutine profile.
All Go services in our monorepo, by default, have profiling enabled, allowing on-demand extraction of goroutine profiles over the network.
Enabling profiling incurs no
overhead since enabling alone does not begin profile capturing.
The overhead of profile capturing is also small and minimized by the sparse collection frequency.



\begin{figure}
    \centering

        \centering
        \begin{tikzpicture}
            \node[style=park, fill=white, draw=white] (header1) {Stack};
            \node[style=park, below of=header1, node distance=1.5em] (gopark) {\mgo{runtime.gopark}};
            \draw [decorate, decoration = {brace, raise=6.5em}] (gopark.north) --  (gopark.south) node[pos=0.5,right=6.5em]{$
                \begin{array}{l}
                     \textrm{\footnotesize Blocked indicator}
                \end{array}
            $};
        
            \node[style=chanop, below of=gopark, node distance=1.5em] (chansend) {\mgo{runtime.chansend}};
            \node[style=chanop, below of=chansend, node distance=1.5em] (chansend1) {\mgo{runtime.chansend1}};
            \draw [decorate, decoration = {brace, raise=6.5em}] (chansend.north) --  (chansend1.south) node[pos=0.5,right=6.5em]{$
                \begin{array}{l}
                     \textrm{\footnotesize Send operation sub-stack}
                \end{array}
            $};
        
            \node[style=caller, below of=chansend1, node distance=1.5em] (sender) {\mgo{server.ComputeCost\$1}};
            \draw [decorate, decoration = {brace, raise=6.5em}] (sender.north) --  (sender.south) node[pos=0.5,right=6.5em]{$
                \begin{array}{l}
                     \textrm{\footnotesize Sender function }
                     \vspace{-0.3em}\\
                     \textrm{\footnotesize (Blocks at Line~\ref{lst:ex:sendmsg})}
                \end{array}
            $};
        
        \end{tikzpicture}
        \caption{Concrete, profile-extracted stack of a goroutine blocked at \texttt{transactions/cost.go:\ref{lst:ex:sendmsg}}.}
        \label{fig:sending-example-stack}
\end{figure}
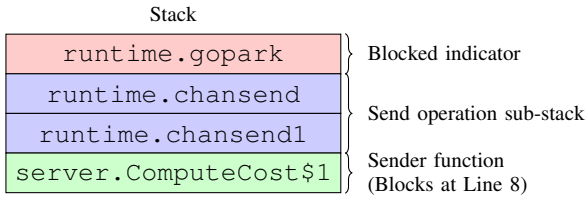

\leakprof{} taps into the goroutine profiles to find goroutine leaks by fetching them periodically (once per day) from every service instance.
While analyzing these profiles, it marks blocked goroutines as candidates for further inspection.
Blocking goroutines exhibit a distinct call stack signature---all blocked goroutines feature \mgo{runtime.gopark} at the top of the call stack\footnote{There is a distinction between a blocked goroutine and one that is not running simply as a result of having been context-switched by the scheduler.}, irrespective of the type of blocking operation (semaphore, IO, garbage collection, channel/select).
Goroutines blocked at channel operations are then recognized by the stack frames right underneath \mgo{runtime.gopark}, as shown in Fig~\ref{fig:sending-example-stack}:


\begin{enumerate}
    \item \mgo{runtime.chansend} and \mgo{runtime.chansend1} indicate a blocking send operation. Receive and select implement similar patterns.
    \item The caller, \mgo{ComputeCost\$1}\footnote{The suffix \mgo[\footnotesize{}]{\$1} denotes the first anonymous function, syntactically, in \mgo[\footnotesize{}]{ComputeCost}}, invoked \mgo{chansend1}, from the \mgo{runtime} library, at \texttt{.../transaction/cost.go:\ref{lst:ex:sendmsg}}, indicating the source location of the send operation.
\end{enumerate}
Thus, every goroutine can be categorized based on what type of channel operation it is blocked on (send/receive/select operation) and further grouped by operation source location.


\textbf{Detection of potential leaks:}
While the number of goroutines blocked on channel operation at a specific location is not definitive proof of a leak in general, our empirical observations favor the notion that an abnormally large concentration of blocked goroutines at a single operation strongly suggests a leak. 
Based on this insight, we have devised heuristics to identify likely false positives effectively. These heuristics are outlined below:
\begin{enumerate}
    \item The count of blocked goroutines originating from the same source location within a process (profile file) is below a specified threshold.
    \label{enum:threshold}
    \item It can be readily demonstrated that the identified blocked candidate goroutine will ultimately become unblocked.
    \label{enum:select}
\end{enumerate}

In the concurrency model of Go, it is inevitable, perhaps even expected, that some goroutines may occasionally block, so blocked goroutines observed via profiling do not necessarily indicate a leak. 
Criterion \ref{enum:threshold} was enforced to reduce the number of spurious reports. 
Only blocking operations exceeding the threshold are marked suspicious for every individual profile.
The threshold is set to 10K blocked goroutines at the same source location in a program; the threshold was determined empirically by starting at a larger number and slowly reducing it as long as the ratio of true positives remained high.
With this 10K threshold, even false positives may sometimes still reveal convoluted patterns leading to congestion that would warrant a redesign.

Criterion \ref{enum:select} attempts to spot identifiable harmless operations. For example, some select statements feature only transiently blocking case arms, e.g., when listening to \texttt{time.Tick}\cite{golangtime} and \texttt{context.Done}\cite{golangctx}. 
Such trivially non-blocking operations are filtered through simple AST-level static analyses.


\textbf{Reporting potential defects:}
The final component of \leakprof{} is the automatic reporting of suspicious operations in goroutine profiles. 
Using our infrastructure for service profiling, \leakprof{} compiles a comprehensive, platform-wide snapshot of goroutine configurations. 
Blocking operations, their underlying stacks, and frequency of occurrence are found and organized by following the outline presented in Section \ref{sec:leakprof-profiles}, with additional structuring as our platform requires.

We have noticed that unusual circumstances, like outages or load imbalances, tend to activate \pds{} in just a few instances. 
We aim to elevate these potential \pd{} reports because of the limited reporting timeframe. 
To achieve this, we conduct a perceived impact evaluation by calculating the \textit{root mean square} (RMS) based on the count of blocked goroutines at a specific blocking source location across profiles from all service instances.
RMS was selected for its capability to effectively highlight suspicious operations within individual instances that exhibit significant clusters of blocked goroutines.

After ordering potential leaks based on their perceived impact, \leakprof{} determines source code ownership and alerts the owners of the top $N$-most impactful blocking locations. 
The project owners may then manually triage these issues. Since manual intervention is still required to address the issue, the report carries relevant information, such as:
\begin{itemize}
    \item the offending operation, source location, and the number of goroutines it blocks,
    \item the representative profile from most blocked goroutines,
    \item the memory footprint over time 
\end{itemize}


\subsection{\leakprof{} overhead}
Throughout its daily runs, on average, \leakprof{} takes about three hours to fetch goroutine profiles from \numSvc{} services deployed as $\approx\kern-0.25em200$K instances running on over a million CPU cores.
Each service has varying instances ranging from tens to thousands, each contributing one goroutine profile.
Individual instances themselves experience statistically insignificant performance overhead during profile collection.
Most of the overhead is incurred in sweeping over all instances running in the system and transferring each profile file over the network.
Analyzing $\approx\kern-0.25em200$K profile files across our platform for suspicious blocking operations takes $<\kern-0.25em1$ minute on average when running on a 48-core Intel Cascadelake CPU with 256 GB DRAM running Linux 3.18.2.
Thus, the leak discovery phase is incredibly lightweight. 
Interfacing with our report routing infrastructure takes an additional $\approx \kern-0.2em3$ minutes.

\section{Findings from \goleak{}}
\label{sec:goleak_findings}

It is impractical to determine the number of unique leaks that \goleak{} prevented post-deployment, as the tool operates on individual developers' machines. 
Tests fail when \pds{} are identified, and developers iteratively address these issues. 
However, we can assess the count of leaks introduced before deploying \goleak{} in our system by retroactively applying the tool across the complete monorepo testing suite. 
This approach enables the identification of weekly increments in newly introduced \pds{}. 
To collect this information, we retrofitted \goleak{} over the preceding 21 weeks leading up to the current deployment date.

\begin{figure}[!t]
    \centering
    \includegraphics[width=\linewidth]{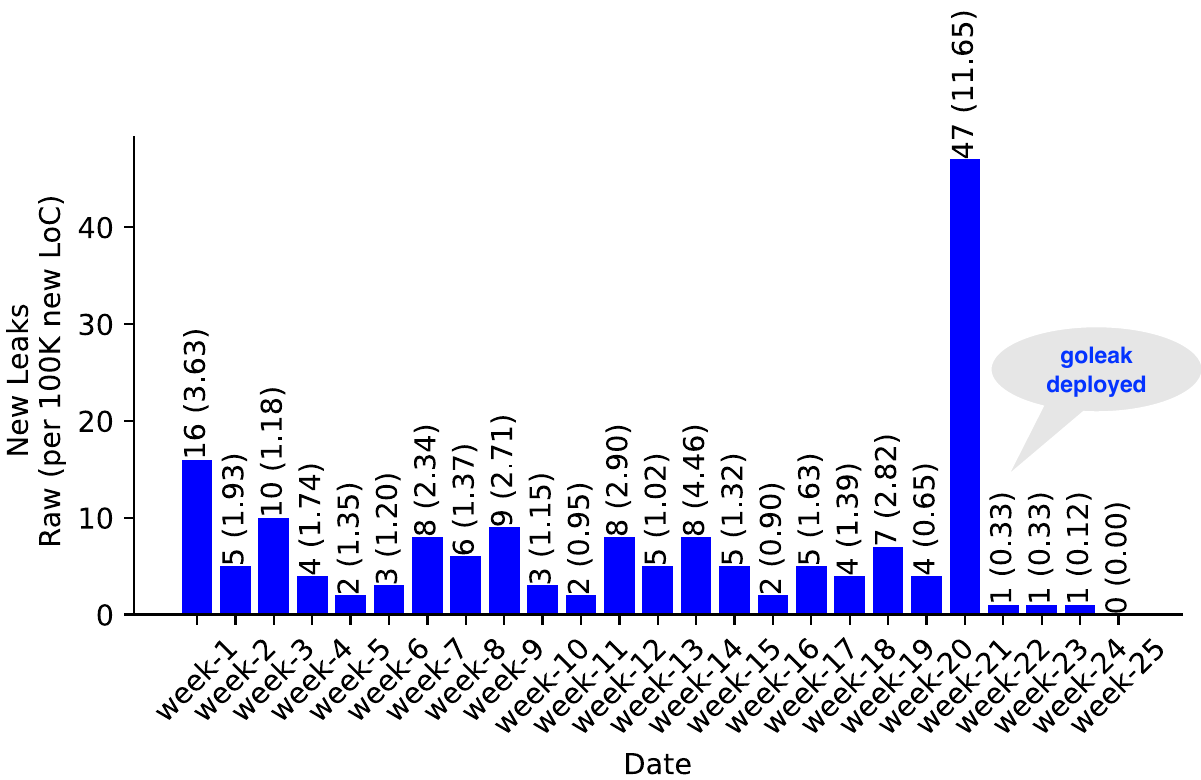}
    \caption{Inflow of new goroutine leaks every week over a 25-week period. The incoming bugs drop to near zero after deploying \goleak{}. The data is collected by backporting the \goleak{} tool.}
    \label{fig:goleakBackptach}
\end{figure}

Figure~\ref{fig:goleakBackptach} plots the number of new \pds{} that were introduced into the system every week.
The x-axis is weeks, and the y-axis shows the raw number of new \pds{} found.
The number in parentheses shows the data normalized by the number of new lines of code added.
The \goleak{} tool was deployed in week 22, after which there was a drastic drop in the number of leaks.
From week one till 21, we see a median of 5 \pds{} introduced per week.
This number is about 1.8 \pds{} per 100K lines of code.
Coincidentally, in week 21, just before deploying \goleak{}, a major project migration brought 47 new \pds{} into the system.
However, immediately after the deployment (weeks 22-24), the number of leaks was not reduced to zero; we saw one new \pd{} added each week because some critical ongoing PRs were landed without fixing the new \pds{} but instead by adding them to the suppression set to be fixed later.
With five new leaks per week, in one year, we expect to see $\approx$260 new leaks to be introduced into the system, which the \goleak{} tool prevented by blocking such PRs.

The initial suppression list size was 1040, of which 857 were related to \pds{}; the remaining were other kinds of runaway goroutines, which \goleak{} found.
Over the past year (weeks 25-75 are not shown since they have near zero new leaks), the number of items in the suppression list has fluctuated as some were fixed and also because some new locations were added to the set.
The current suppression number is 1056, which is 16 new locations compared to the start. 
This is significantly lower than the 260 leaks that could have been introduced had \goleak{} not been deployed.

We classified all \pds{} found by \goleak{} into three categories grouped by their unique source locations (two defects arising from the same source location are counted only once): 
\begin{itemize}
    \item blocked on sending over a channel (15\%),
    \item blocked on receiving on a channel (40\%), and
    \item blocked in the select statement (45\%)
\end{itemize}


The imbalance in the number of leaks introduced in channel receive (40\%) vs. channel send (15\%) was surprising.
In fact, it was counter-intuitive because we were expecting more goroutines to be blocked in channel send operations akin to the example we showed in Listing~\ref{lst:example} due to a premature return of channel receivers.
We investigated over 500 randomly sampled preexisting leaks and found out that there are a large number of \pds{} in channel receive arising from the following two common leaky coding patterns followed by Go developers.

\subsection{Patterns leading to \pds{} in channel receive.}
\label{sec:leaky-patterns-test}

Channel receive \pds{} are predominantly caused by non-terminating timers ($44\%$) and loop-ranges over channels that were never closed
($42\%$), with only the remaining $14\%$ falling in the other categories.

\subsubsection{Loop iteration over unclosed channels}
Go allows the \mgo{for} loop to range over a channel.
The loop extracts one item each from the channel in the order it was inserted.
Such loops continue iteration until the channel is closed by calling the \mgo{close} built-in operation with it as an argument. 
Typically, the channel is closed by a partner goroutine.
However, if \mgo{close} is not called, the receiver loop blocks forever after dequeuing the last item from the channel.
It is also possible for multiple goroutines to iterate over the same channel, and Listing~\ref{lst:unclosedchan} shows the case of a single producer with multiple consumers.
Line~\ref{pc:make} creates a shared, unbuffered channel \texttt{ch}.
At lines~\ref{pc:producer}-\ref{pc:producerend}, we have the concurrent producer, which inserts items into the channel.
Before that, at lines~\ref{pc:consumer}-\ref{pc:consumerend}, a variable number of goroutines, equal to \mgo{workers}, is created, each looping over items inserted into the channel. 
After all items are inserted, the channel should be closed so all receivers can exit the channel loops (Line~\ref{pc:loop}) and terminate.
However, the absence of \mgo{close(ch)} leads all the consumers to block indefinitely.

\begin{figure}[!t]
        \centering
        \begin{lstlisting}[language=go,label={lst:unclosedchan},caption={Iterating over unclosed channel.}]
func producerConsumer(items []Item, workers int) {
@\label{pc:make}@  ch := make(chan Item)
  // Multiple concurrent consumers.
  for i := 0; i < workers; i++ {
@\label{pc:consumer}@    go func () { 
@\label{pc:loop}@      for item := range ch {
        // process item
        ...
      }
@\label{pc:consumerend}@    }()
  }

  // Concurrent producer. 
@\label{pc:producer}@  for item := range items {
@\label{pc:send}@    ch <- item
@\label{pc:producerend}@  }  
}        
        \end{lstlisting}
\end{figure}

\subsubsection{Infinite receive loops with timers}
Developers often introduce operations performed on periodic heartbeats. 
These operations involve producing metrics or logs, or flushing in-memory data to a persistent store, or evicting a cache on a timer, to name a few.
Such tasks are often delegated to goroutines created upon the invocation of some higher-level API.
Listing~\ref{lst:timerloop} shows an example where the function \mgo{statsReporter} launches a goroutine that periodically logs some metrics in an infinite loop.
The call to \mgo{time.After} on Line~\ref{timer:after} is a channel receive operation, which blocks until there is a message on the channel. 
The channel returned by \mgo{time.After} is sent to after a \mgo{reporterPeriod} timer expires, which is an idiomatic way to stall goroutine execution for specific amounts of time.
While this pattern does not strictly constitute a \pd{}, since the goroutine periodically wakes up and goes back to sleep, it still constitutes an anti-pattern since there is no control over the lifetime of the goroutine launched by \mgo{statsReporter}.
A better approach to writing such code would be to place \mgo{<-time.After} as a case in a select statement, where the other case waits for a termination message that can be issued by the caller.

\begin{figure}[!t]
        \centering
        \begin{lstlisting}[language=go,label={lst:timerloop},caption={Infinite receive loops with timers.}]
func statsReporter() {
  go func() {
    for {
@\label{timer:after}@      <-time.After(reporterPeriod)
      LogMetric()
    }
  }()
}
        \end{lstlisting}
\end{figure}

\subsection{Patterns leading to \pds{} in channel send.}

Among the \pds{} blocked on send operations, 57\% were due to the receiver prematurely returning due to timeout or errors; 11\% were due to library API callers creating the sender but either not creating or incorrectly using the receiver; 29\% were due to other causes involving complex state machine; 3\% were due to the double send pattern described below.

\subsubsection{The double send issue} Listing~\ref{lst:doublesend} shows the double send issue.
The \mgo{sender} and the \mgo{receiver} functions share a common channel \mgo{ch} for communicating \mgo{interface} objects.
The sender attempts to create an item (Line~\ref{ds:errHappens}); if the creation fails, it sends a \mgo{nil} value (Line~\ref{ds:nilsend}) on the channel; otherwise, she sends the item on the channel (Line~\ref{ds:senddouble}). 
However, the developer forgot to add a \mgo{return} statement immediately after Line~\ref{ds:nilsend}, eventually leading to the second send operation.
Since the receiver only accepts one message on Line~\ref{ds:rcv}, whenever \texttt{createItem} returns an error, the second send operation causes a \pd{}. 

\begin{figure}[!t]
        \centering
        \begin{lstlisting}[language=go,label={lst:doublesend},caption={Double send \pd{}.}]
func sender(ch chan interface{}) {
@\label{ds:errHappens}@ item, err := createItem()
 if err != nil {
    // error occurred, send nil
@\label{ds:nilsend}@    ch <- nil
  }
@\label{ds:senddouble}@  ch <- item
}

func receiver(ch chan interface{}) {
@\label{ds:rcv}@  item := <- ch
  // process item  
}        
        \end{lstlisting}
        \label{fig:doublesend}
\end{figure}

\subsection{Patterns leading to \pds{} in select statements.}

\PDs{} resulting from select statements consist of method contract violation ($86.16\%$), infinite \mgo{for} loops with no escape hatch (execution paths leading to \mgo{return}/\mgo{break} statements, $7.7\%$), and select statements without cases, which always block indefinitely ($6.16\%$).
Select statements with a single arm are desugared at compile time into individual send or receive operations, so they fall under the purview of the previous categories.

\subsubsection{Method contract violation} Listing~\ref{lst:method-contract} showcases a \pd{} caused by a method contract violation.
The type \mgo{Worker} (Line~\ref{ds:worker-type}) embeds two channel fields: \mgo{ch} and \mgo{done}.
Invoking the \mgo{Start} (Line~\ref{ds:start-func}) method creates a goroutine (Line~\ref{ds:start}), which continuously listens for messages on the embedded channels. 
The listener terminates if it either receives a message on \mgo{w.done}, or after \mgo{w.done} is closed. 
Since the \mgo{Stop} (Line~\ref{ds:stop-func}) method closes the \mgo{done} channel, the lifetime of the listener goroutine is therefore bounded by the invocation of \mgo{Stop}. 
As a consequence, the \mgo{Start} and \mgo{Stop} methods establish an implicit contract, according to which every invocation of \mgo{Start} should always eventually be followed by an invocation of \mgo{Stop} to allow any listener goroutines to terminate. 
Omitting an invocation of \mgo{Stop} during the lifetime of a \mgo{Worker} leads to the \pd{} of the listener.

Method contract violation \pds{} similar to the one exemplified in Listing \ref{lst:method-contract} are the most common in practice ($58.47\%$ of all select \pds{}), but this class of \pds{} has several variations. 
These include replacing the \mgo{done} channel with other standard library features e. g., \mgo{context.Context} ($16.93\%$), or
blocking at a \mgo{select} statement outside a \mgo{for} loop ($27.7\%$).

\begin{figure}
        \centering
        \begin{lstlisting}[language=go,label={lst:method-contract},caption={Partial deadlock caused by method contract violation.}]
@\label{ds:worker-type}@type Worker struct {
  ch chan any
  done chan any
}

@\label{ds:start-func}@ func (w Worker) Start() {
  @\label{ds:start}@ go func() {
    for {
      @\label{ds:select}@select {
        @\label{ds:case-normal}@case <-w.ch: // Normal workflow
        @\label{ds:case-done}@case <-w.done:
          return // Shut down
      } // end select
    } // end for
  }() // end go
}

@\label{ds:stop-func}@func (w Worker) Stop() {
    // Allows goroutine at line @\ref{ds:start}@ to exit
    close(w.done)
}

func foo() {
  w := Worker{ ch: make(chan any); done: make(chan any) }
  w.Start()
  // Exits without calling 'Stop'
  return
}        
        \end{lstlisting}
        \label{fig:doublesend}
\end{figure}

\subsection{Magnitude of leaks across the monorepo}
After running all \goNumTests{} tests, we classified all non-terminated goroutines by their blocking type.
This analysis is intended to provide a magnitude of the frequency of leaks caused by different reasons; hence, we do not deduplicate them by unique source location here.
Since \goleak{} detects any lingering goroutine, we provide the frequency of other types of runaway goroutines in addition to channel-related leaks.
Table~\ref{tab:leakclassification} shows the summary.
\emph{Message-passing is the cause for over 80\% of all non-terminated goroutines.}
Moreover, blocking in select statements is the single largest ($>50\%$) cause, indicating the propensity of developers to misuse select statements.
Blocking in channel receive is the second largest cause ($\approx$32\%), significantly more than blocking in channel send ($\approx$1.73\%). 
The reason for the bias in blocking over receives is already explained in the previous subsection. 
Sending or receiving on a \mgo{nil} channel or waiting on a select statement with no case arms guarantee a \pd{}, and hence, they are shown separately.

\begin{table}[!t]
\small
    \centering
    \begin{tabular}{|r|r|r|}
        \hline
    \textbf{Type} & \textbf{Count} & \textbf{Percentage} \\ \hline
        chan receive (non-nil chan) & 46K  & 32\% \\
        chan receive (nil chan) & 14 & 0.01\%  \\ \hline
        chan send (non-nil chan) &  2.5K & 1.73\%\\
        chan send (nil chan) &  5 & 0.003\% \\ \hline
        select ($>\kern-0.2em0$ cases) & 75K & 51\%\\ 
        select (0 cases) &  10 & 0.007\%\\ \hline
        IO wait & 9K & 6.4\% \\
        System call & 6.4K & 4.4\%  \\
        Sleep & 5.5K & 3.8\% \\
        Running/Runnable & 407 & 0.27\% \\
        Condition Wait & 46 & 0.03\% \\
        Semaphore Acquire & 138 & 0.09\%\\ \hline
        \textbf{Total} & 164K & 100\% \\
        \hline
    \end{tabular}
    \caption{Classification of blocking types over all non-terminating goroutines after running all tests.}
    \label{tab:leakclassification}
\end{table}

\section{Findings from \leakprof{}} 
\label{sec:lekprof_findings}
During a one-year deployment and observation period, \leakprof{} reported 33 suspected defects from production executions, of which the developers acknowledged 24. 
Of these, 21 issues were resolved with a fix. 

\begin{figure}
    \centering
    \includegraphics[width=0.9\linewidth]{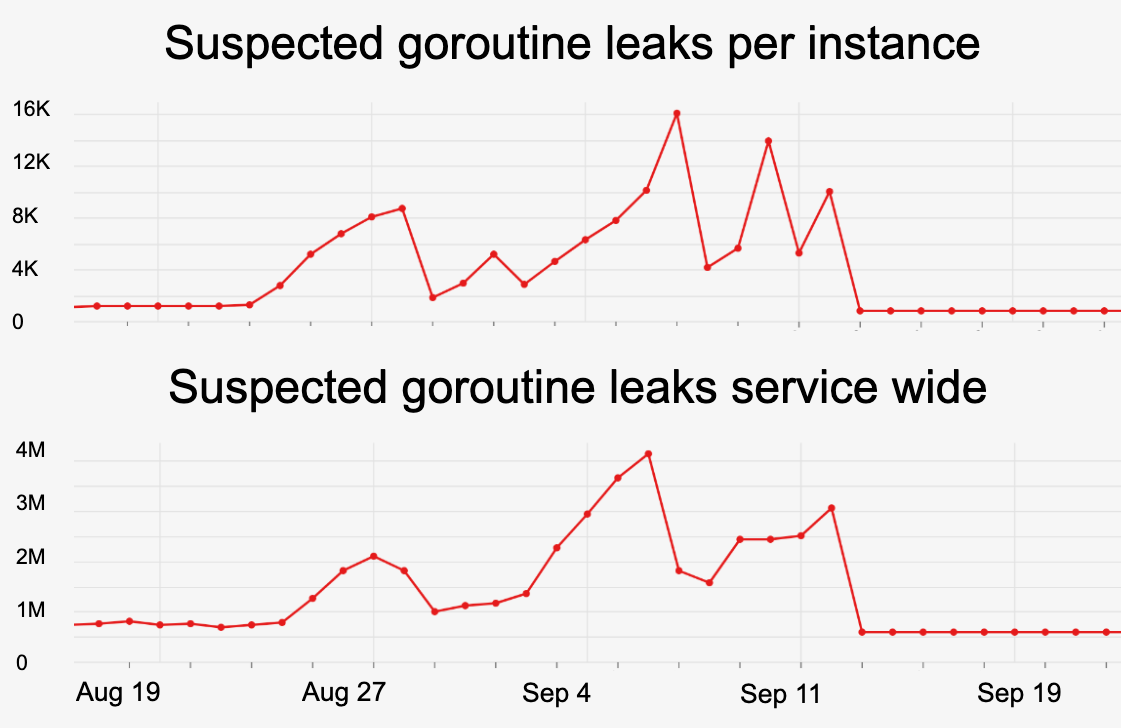}
    \caption{Example blocked goroutine footprint for a leaky service: for representative instance (top), and across the entire fleet (bottom).}
    \label{fig:resource-insights}
\end{figure}

Fig~\ref{fig:resource-insights} presents how \leakprof{} offers observability into channel-based goroutine leaks in production services.
As evidenced from the time-series data, a new goroutine leak bug was introduced into production, leading to an excessive number of goroutines blocked on a channel operation.
The leak caused $\approx~3$ million blocked goroutines across 800 instances of the service (bottom half of the figure), with one representative instance spiking to up to $16K$ blocked goroutines at a single source location (top half of the figure).
Fortunately, after the number of blocked goroutines exceeded the threshold, \leakprof{}~allowed us to intercept and address the leak promptly.

We document in more detail the impact of both goroutine leaks and deployed fixes on several affected services (denoted via pseudonyms $S_i$) in Table \ref{table:mem-reductions}. 
In the \textbf{Service wide peak utilization} columns, we show the peak memory footprint for the given service (accumulated over all deployed instances) before and after the fix. 
In the \textbf{Capacity} columns, where applicable, we document the memory capacity reductions for individual service instances after the deployment of the fix, as deemed appropriate by service owners, quantifying the real cost savings. 
Notably, for services $S_7$ and $S_9$ we discovered that instances were significantly overprovisioned relative to actual consumption even before the fix. This suggests that manual or automatic scaling was employed to accommodate the ever-increasing memory requirements of affected services over protracted periods, with \pds{} as a contributing factor.

\begin{table}
\small
    \centering
    \begin{tabular}{|l||r|r||r|r|}
        \hline
        \multirow{2}{*}{
        \hspace{-1.3em}
        $\begin{array}{c}
        \textbf{Service} \\
        \textbf{(\#instances)}
        \end{array}$} 
        \hspace{-1.3em} &
        \multicolumn{2}{|c||}{
        $\begin{array}{c}
        \textbf{Service wide} \\
        \textbf{peak utilization (GB)}
        \end{array}$} &
        \multicolumn{2}{|c|}{
        $\begin{array}{c}
            \textrm{\textbf{Capacity (GB)}} \\
            \textrm{\textbf{per instance}}
        \end{array}$}
        \\
        \cline{2-5}
        & \textbf{Before} &
        \hspace{-1em}
        $\begin{array}{c}
        \textrm{\textbf{After}}  \\
         \textrm{\textbf{(\% saved)}}
        \end{array}$
        \hspace{-1em}
        & \textbf{Before} &
        \hspace{-1em}
        $\begin{array}{c}
        \textrm{\textbf{After}}  \\
         \textrm{\textbf{(\% saved)}}
        \end{array}$
        \hspace{-1em} 
        \\
        \hline
         $S_1$ (5854) & 28K & 13K (53\%) & 4 & - (0\%) \\
        \hline
        $S_2$ (612) & 310 & 290 (9\%) & 5 & 4 (20\%) \\
         \hline
         $S_3$ (199) & 317 & 182 (42\%) & 4 & 3 (25\%) \\
        \hline
        $S_4$ (120) & 116 & 72 (37\%) & 6 & 4 (25\%)\\
        \hline
        $S_5$ (72) & 650 & 347 (46\%) & 17 & - (0\%) \\
        \hline
        $S_6$ (66) & 112 & 36 (67\%) & 4 & 3 (25\%) \\
        \hline
        $S_7$ (64) & 83 & 63 (29\%) & 43.5 & 3 (92\%) \\
         \hline
         $S_8$ (19) & 35 & 29 (17\%) & 8 & 6 (25\%)\\
         \hline
         $S_9$ (18) & 30 & 6.5 (78\%) & 32 & 8 (75\%) \\
        \hline
        $S_{10}$ (10) & 19 & 15 (21\%) & 4 & 3 (25\%)\\
         \hline
         $S_{11}$ (9) & 4.5 & 3.3 (26\%) & 8 & - (0\%) \\
         \hline
         $S_{12}$ (6) & 9.6 & 4.2 (56\%) & 4 & - (0\%) \\
         \hline
        $S_{13}$ (6) & 7.5 & 2 (73\%) & 4 & 3 (25\%) \\
        \hline
    \end{tabular}
    \caption{Impact of fixes on peak memory footprint, and reduced capacity for individual instances of a service after the fix.}
    \label{table:mem-reductions}
\end{table}

\subsection{Partial deadlock patterns from production}
\label{sec:leaky-patterns-prod}

\begin{figure}[!t]
\begin{minipage}[t]{\linewidth}

        \centering
        \begin{lstlisting}[language=go,escapechar=|,label={fig:premature-return-leak},caption={Premature function return.}]
ch := make(chan T)
go func() {
    ch <- t |\label{lin:premature-return-send}|
}()
if ... {
    return |\label{lin:premature-return-return}|
}
<-ch
        \end{lstlisting}
\end{minipage}
\hfill
\begin{minipage}[t]{\linewidth}
        \centering
        \begin{lstlisting}[language=go,escapechar=|,
        label={fig:timeout-leak},
        caption={The timeout leak.}
        ]
func Handler(ctx context.Context, ...) {
  ch := make(chan Item)
  go func() {
    ch <- item |\label{lin:timeout-send}|
  }()
  select {
  case item := <-ch: // Process item
  case <-ctx.Done(): return|\label{lin:timeout-done}|
  }
  ...
}
        \end{lstlisting}
    
\end{minipage}
\hfill
\begin{minipage}[t]{1\linewidth}
        \centering
        \begin{lstlisting}[language=go,escapechar=|,
        caption={The NCast leak.},
        label={fig:ncast-leak}
        ]
ch := make(chan T)
for _, item := range items {
    go func(t T) {
        ch <- t |\label{lin:ncast-send}|
    }(item)
}
<-ch |\label{lin:ncast-recv}|
        \end{lstlisting}
\end{minipage}
\end{figure}

In this section, we share some of the common patterns of \pds{} we observed in alerts issued by \leakprof{}, simplified such that they capture the essence of the leaks while giving enough context for the reader to appreciate the impact.

\subsubsection{Premature Function Return Leak}
\label{sec:premature-return}

\textbf{(4 reports)} Listing~\ref{fig:premature-return-leak} illustrates an example of misuse of conditional branches, such that the communicating behavior is uneven between execution paths. If the parent goroutine follows the \mgo{true} branch of the \mgo{if} statement, it will prematurely return (Line~\ref{lin:premature-return-return}) without synchronizing on channel \mgo{ch}. Because \mgo{ch} is unbuffered and its scope is limited to the listed snippet, the sender will be permanently stuck (Line~\ref{lin:premature-return-send}). The simplest solution is to give \mgo{ch} a buffer of size one, unblocking the send unconditionally.





\subsubsection{The Timeout Leak}

\textbf{(5 reports)} while this pattern may be framed as a special case of the Premature Return (Section~\ref{sec:premature-return}) pattern, its sheer ubiquity\cite{DBLP:conf/asplos/LiuZQCS21} elevates it to its own entry. 
Listing~\ref{fig:timeout-leak} shows an example of a function \texttt{Handler}, which takes a \texttt{context} argument.
\texttt{Context}s~\cite{golangctx} in Go carry request-scoped values such as deadlines and cancellation signals across API boundaries and between processes.
The \texttt{Handler} implementation creates a channel \mgo{ch} and then creates a goroutine which sends a message on \mgo{ch}.
Using a select statement, the \texttt{Handler} then waits for either a message on \mgo{ch} or the context to be canceled.
The \texttt{Handler} can be terminated by its invoker by canceling the context, which sends a signal on the \texttt{ctx.Done()} channel. 
If the context is canceled before the  \texttt{item}  is received from the channel, the \texttt{Handler} returns, and the anonymous goroutine blocks forever in sending the \texttt{item} on \mgo{ch} with no matching receiver.
The simplest solution is to make the channel non-blocking by increasing the capacity of \mgo{ch} to one so the sender does not block whether or not the receiver is present.

\subsubsection{The NCast Leak}

\textbf{(4 reports)} this \pd{} is caused when one communicating party performs more sends/receives than its partner.
Listing~\ref{fig:ncast-leak} illustrates such a case, where   \mgo{len(items)} number of messages are sent\footnote{Assuming \lstinline[identifierstyle=\footnotesize\ttfamily{},basicstyle=\footnotesize\ttfamily{}]{items} is not empty} but only one channel receive is executed on an unbuffered channel.
The developer intends to wait for the first returned result and ignore the rest.
The first goroutine to send on the channel finds a pairing receiver and unblocks;
however, all other sender goroutines (Line~\ref{lin:ncast-recv}) find no receiver and block permanently. 
A remedy to fix this leak while maintaining the original semantics would be to adjust the channel's capacity to \mgo{len(items)} such that all the send operations are guaranteed to unblock.

\subsubsection{Remaining reports}

For illustrative purposes, the remaining defects detected by \leakprof{} fall under the following categories: \textit{double send \pds{}} \textbf{(2 reports)}, \textit{channel iteration without close} \textbf{(2 reports)}, \textit{method contract violation} \textbf{(1 report)}, and \textbf{6 reports} outside any of these categories.

\section{Conclusions and Future work}
\label{sec:conclusion}

Go programming language makes it easy to write concurrent code, but does not protect against common pitfalls of concurrent programming; in fact, supporting both shared memory and message-passing in Go makes it susceptible to a multitude of concurrency bugs: data races, atomicity violations, and deadlocks, to name a few.

The continuous development in large-scale industrial systems leads to a large and constantly shifting surface area for errors.
In practice, tests are the first line of defense against faulty code, but the Go testing framework does not automatically guard against \pds{}. 
We account for this deficiency by leveraging test instrumentation. 
Even so, there may still be inputs, path conditions, and interleavings of complex interacting concurrent systems without proper test coverage, potentially allowing \pds{} to still sneak into production and linger for a long time, which motivates the deployment of runtime monitoring systems.
While neither sound nor complete, our approach is effective in detecting suspicious high concentrations of blocked goroutine in active services.
Combining both techniques has allowed us to address several taxing \pds{} effectively.

The patterns identified in Sections \ref{sec:leaky-patterns-prod} and \ref{sec:leaky-patterns-test} also give ample insight into the most common types of \pds{} introduced by developers. 
This ultimately points us toward targeted and effective static techniques for preventing \pds{}. 
For our initial efforts, in the hopes of preventing \pds{} similar to those in Listing \ref{lst:unclosedchan}, we have already designed a \textit{range linter} that reports whether local, lexically scoped channels used with the \mgo{range} construct may never be closed.
The field is open for many such lightweight automation that can employ both static and dynamic analysis and target high-value defects with low overhead for industry-scale use in mission-critical systems.
\newpage
\bibliographystyle{ieeetran}
\bibliography{ref}
\end{document}